\documentclass[aps,pre,twocolumn,showpacs,floatfix,superscriptaddress,nofootinbib]{revtex4}
\usepackage{graphicx}
\usepackage{amsmath}

\usepackage[usenames,dvipsnames]{color}

\newcommand {\be} {\begin{equation}}
\newcommand {\ee} {\end{equation}}
\newcommand {\Be}{\begin{eqnarray*}}
\newcommand {\Ee} {\end{eqnarray*}}
\newcommand {\bey} {\begin{eqnarray}}
\newcommand {\eey} {\end{eqnarray}}
\newcommand{\bit}{\begin{itemize}}      
\newcommand{\eit}{\end{itemize}}
\newcommand{\bfl}{\begin{flusleft}}
\newcommand{\efl}{\end{flusleft}}
\newcommand{\bfr}{\begin{flushright}}
\newcommand{\ec}{\end{center}}
\newcommand{\ben}{\begin{enumerate}}    
\newcommand{\een}{\end{enumerate}}

\newcommand{\comment}[1]{}

\begin{document} 

\title{Boundary-induced instabilities in coupled oscillators}

\author{Stefano Iubini}
\affiliation{Dipartimento di Fisica e Astronomia and CSDC, Universit\`a di Firenze, 
via G. Sansone 1 I-50019, Sesto Fiorentino, Italy}
\affiliation{Istituto Nazionale di Fisica Nucleare, Sezione di Firenze,  
via G. Sansone 1 I-50019, Sesto Fiorentino, Italy}
\affiliation{Consiglio Nazionale delle Ricerche, Istituto dei Sistemi Complessi, 
via Madonna del Piano 10, I-50019 Sesto Fiorentino, Italy}

\author{Stefano Lepri}
\email{stefano.lepri@isc.cnr.it}
\affiliation{Consiglio Nazionale delle Ricerche, Istituto dei Sistemi Complessi, 
via Madonna del Piano 10, I-50019 Sesto Fiorentino, Italy}
\affiliation{Istituto Nazionale di Fisica Nucleare, Sezione di Firenze,  
via G. Sansone 1 I-50019, Sesto Fiorentino, Italy}

\author{Roberto Livi}
\affiliation{Dipartimento di Fisica e Astronomia and CSDC, Universit\`a di Firenze, 
via G. Sansone 1 I-50019, Sesto Fiorentino, Italy}
\affiliation{Istituto Nazionale di Fisica Nucleare, Sezione di Firenze,  
via G. Sansone 1 I-50019, Sesto Fiorentino, Italy}

\author{Antonio Politi}
\affiliation{Institute for Complex Systems and Mathematical Biology \& SUPA
University of Aberdeen, Aberdeen AB24 3UE, United Kingdom}

\begin{abstract}
A novel class of nonequilibrium phase-transitions at zero temperature 
is found in chains of nonlinear oscillators.
For two paradigmatic systems, the Hamiltonian XY model and the discrete nonlinear Schr\"odinger 
equation, we find that the application of boundary forces  
induces two synchronized phases, separated by a non-trivial interfacial region
where the kinetic temperature is finite. Dynamics in such supercritical state 
displays anomalous chaotic properties whereby some 
observables are non-extensive and transport is superdiffusive. At finite temperatures, 
the transition is smoothed, but the temperature profile is still non-monotonous.
\end{abstract}

\pacs{05.45.Xt 05.70.Ln 05.60.-k}


\maketitle

The characterization of steady-states is a widely investigated problem within 
non--equilibrium statistical mechanics \cite{Bertini07}, since it provides the basis for
understanding a large variety of  phenomena, including transport 
processes, pattern formation and the  dynamics of living systems.
In a nutshell, the simplest setup amounts to determining the currents that emerge
as a result of the application of an external force, either across the system,
as for electric currents, or at the boundaries, as in heat 
conduction~\cite{LLP03,DHARREV,Basile08}. 
Anyway, it is quite a nontrivial task to be accomplished,   
even when the departure from equilibrium is minimal and one can rely
on the Green-Kubo formalism for establishing a connection between the microscopic
and the hydrodynamic descriptions. For instance, this is testified
by the discrepancy that still persists, after many years of careful studies,
between the most advanced theories of heat conduction and some numerical simulations.
The level of difficulty typically increases when one considers coupled transport
\cite{Gillan85,Mejia2001,Larralde03,Saito2010,Basko2011,Iubini2012,DeRoeck2013}
(i.e. when two or more currents coexist, such as heat and electric ones in 
thermo-electric effects) or, even worse, far-from-equilibrium. 
This is why most of the theoretical studies concentrate on 
stochastic models, where fluctuations can be easily controlled,
although they lack a truly microscopic justification.
This approach proved, nevertheless, very effective, since it has allowed
discovering non-equilibrium transitions, such as those exhibited by TASEP-like models,
that have been used to describe translation of proteins, or traffic flows \cite{Derrida1998}.

In this Letter we describe a novel class of boundary-induced transitions 
for two models that are typically used as test beds for a wide range of physical
phenomena: the so-called Hamiltonian XY (or rotor) 
model~\cite{Escande1994,Giardina99,Gendelman2000,Yang2005} subject to an applied
mechanical torque and the Discrete NonLinear Schr\"odinger (DNLS) 
equation ~\cite{Eilbeck1985,Eilbeck2003,Kevrekidis} under a gradient of the chemical potential.
This type of qualitative change of the dynamics results from the joint effect 
of thermal and mechanical forces. It can be interpreted as a desynchronization phenomenon
in a spatially-extended dynamical system,
whereby mutual entrainment of oscillators' phases is abruptly destroyed.
As a result of such unlocking, a regime characterized by phase-coexistence 
sets in where, although the chain is attached to two zero-temperature thermostats,
an interfacial region is spontaneously created, where the oscillators have a finite 
kinetic temperature. Such a state can neither be predicted within a linear-response type of
theory, nor traced back to some underlying equilibrium transition. Even more remarkably, it constitutes 
an example of a highly inhomogeneous, unusual chaotic regime. Indeed, we will 
show that there dynamical invariants have non-standard dependence on the 
system size, as the fractal dimension is extensive while the Kolmogorov-Sinai (KS) entropy
is not.

Studies of unlocking transitions have been previously performed in purely dissipative chains of 
phase-oscillators~\cite{Kopell1990}, where, however, the absence of conservation laws 
prevents the onset of hydrodynamic regimes such as those herein described.
The effect of external forces on the Hamiltonian XY model have been previously addressed 
only in Ref.~\cite{Iacobucci2011} (see also Ref.~\cite{Eleftheriou2005}).
Boundary-induced transitions are also known to exist for other classes nonequilibrium models 
like stochastic lattice gases \cite{Krug1991}. In the present case however
the (zero-temperature) non-equilibrium transition is of purely dynamical origin.

{\bf Hamiltonian XY model}. The model consists of a chain of $N$ rotors whose phases 
$q_n$ evolve according to the equations
\begin{eqnarray}
  \dot{p}_n &=& \sin(q_{n+1}-q_n) - \sin(q_n-q_{n-1})+\\
 && +\left(\delta_{1,n}+\delta_{N,n}\right)[ \gamma (F_n-p_n) + \sqrt{2\gamma T}\,\eta_n]\nonumber
 \label{xymodel}
\end{eqnarray}
where $ p_n =  \dot{q}_n$,  $F_n$ denotes a torque applied to the chain boundaries, $\gamma$ is 
the coupling strength with two external baths and $\eta_n$ is a Gaussian white 
noise with unit variance.
Even though the two heat baths are assumed to have the same temperature $T$, one expects
(coupled) momentum and energy currents will flow through the lattice. The momentum
(angular velocity) flux is defined as $j^p_n=\langle \sin(q_{n+1}-q_n)\rangle$, 
while the energy flux is $j^e_n= \langle (p_n+p_{n+1})\sin(q_{n+1} - q_n)\rangle$~\cite{Spohn2013}
(here and in the following, angular brackets denote a time average). Further useful 
observables are: the average angular frequency 
$\omega_n =\langle p_n \rangle$ of the $n$th oscillator and the kinetic temperature
$T_n=\langle (p_n-\omega_n)^2\rangle$ (notice that a correct definition
requires subtracting the average drift).

\begin{figure}[ht]
\begin{center}
\includegraphics[width=0.45\textwidth,clip]{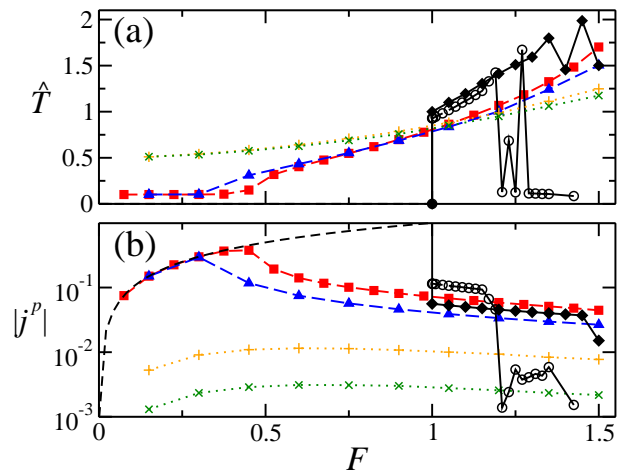}
\caption{(Color online) Phase diagrams of the XY model. 
(a) Maximal kinetic temperature $\hat T$ 
versus $F$; $T = 0$: open circles ($N=200$) and diamonds ($N=6400$);
$T = 0.1$: squares ($N=200$) and triangles ($N=800$); $T = 0.5$: plusses
($N=200$) and crosses ($N=800$). (b) Momentum flux  $j^p$ vs. $F$ for the same
temperatures and symbols of panel (a); the black dashed  line corresponds to
$j^p= F$ for $F<F_c = 1$.
}
\label{F.phasediag}
\end{center}
\end{figure}

We first discuss the $T=0$ case. As long as $F\equiv (F_1-F_N)/2 < F_c=1/\gamma$, 
the ground state is a twisted fully-sinchronized state, whereby each element rotates 
with the same frequency $\omega_n = (F_1+F_N)/2$ and constant phase gradient. 
Here, $T_n=0$ throughout the whole lattice.
For $F>F_c$ the fully sinchronized state turns into a chaotic asynchronous dynamics.
All numerical simulations hereafter reported have been performed with $\gamma = 1$ 
and with  $F_1=-F_N=F$ (that amounts to fix $\omega_n = 0$ below threshold). 
As shown in Fig.~\ref{F.phasediag}a, the maximum value $\hat T$ of $T_n$ along
the lattice suddenly jumps to a finite value at $F=F_c$, indicating the presence of a first-order
nonequilibrium transition. In fact, although the energy flux $j^e$ vanishes (both heat baths
operate at zero temperature), the momentum current $j^p$ is different from zero and undergoes
a substantial drop above the transition point (see Fig.~\ref{F.phasediag}b).

\begin{figure}[htbp]
\begin{minipage}{0.55\linewidth}
\centering
\includegraphics[width=\textwidth,clip]{figure2a}
\end{minipage}%
\begin{minipage}{0.45\linewidth}
\centering
\includegraphics[width=\textwidth,clip]{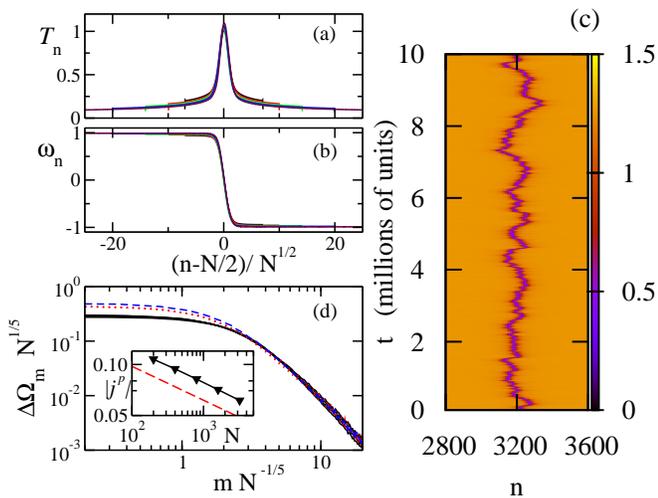}
\end{minipage}
\caption{(Color online) Stationary behaviour of the rotors chain for $T=0$.
Time averaged spatial profiles of  temperature (a) and frequency (b) for $F=1.05$ and
 $N=$200, 400, 800, 1600, 3200  (the averaging time is $t=10^7$).
The spatial direction is rescaled  in order to  obtain a data collapse in the central region.
(c) Spatio-temporal profile of $|\Omega_n|$ for $F=1.3$ and $N=6400$ (see text for details).
(d) Local frequency difference $\Delta\Omega_m$. Both axes are suitably rescaled to
obtain a data collapse for different system sizes:
$F=$1.05 and $N=$400, 800, 1600, 3200 (black lines); 
$F=$1.2 and $N=$1600, 3200, 6400 (dotted red lines);
same parameters of panel (c) (dashed blue line).
The inset shows the dependence on $N$ the momentum flux $j^p$ for $F=1.05$ (black triangles): the red dashed line
has slope -1/5.}  
\label{F.profiles}
\end{figure}

\begin{figure}[ht]
\begin{center}
\includegraphics[width=0.4\textwidth,clip]{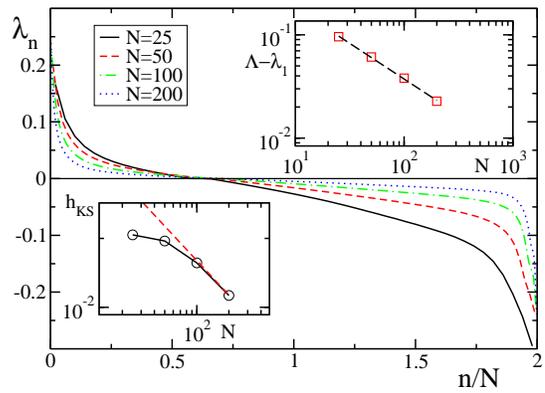}
\caption{(Color online) Lyapunov spectra of the rotors chain for $F=1.05$ and increasing sizes $N$.
Upper inset: dependence  of the maximal Lyapunov exponent 
$\lambda_1$  on $N$: the dashed line corresponds to the fit 
$\lambda_1 = \Lambda + aN^{-0.66\pm 0.04}$ with $\Lambda=0.262$. Lower inset:
Kolmogorov-Sinai  entropy versus $N$; the dashed line is the power-law $N^{-1/2}$.
Total integration time $t=2\times 10^6$, time step $0.01$ time units, Gram-Schmidt 
orthogonalization is applied every $20$ time steps.
}
\label{F.lyspectr}
\end{center}
\end{figure}

A more detailed characterization of the supercritical phase is reported in 
Fig.s~\ref{F.profiles} and \ref{F.lyspectr}.  
The temperature profiles above threshold ($F=1.05$) are shown in Fig.~\ref{F.profiles}a
for different values of $N$, after shifting the origin in the middle of the chain and 
rescaling the spatial position 
by $\sqrt{N}$. The nice overlap has two implications: (i) the maximal temperature $\hat T$ 
remains finite even in the thermodynamic limit and can, accordingly, be considered as 
an appropriate order parameter for this nonequilibrium transition; (ii) $T_n$
is significantly different from zero only in a small central region, whose relative width 
scales as $N^{-1/2}$. Since the temperature is a macroscopic concept, it is legitimate to 
ask whether one can truly interpret $T_n$ as a genuine thermodynamic temperature. 
A preliminary positive answer can be given by noticing that $T_n$ does not vary significantly
over a diverging number ($\approx \sqrt{N}$) of sites.

Additional information can be obtained by looking at the profile of the average angular
frequency $\omega_n$ for $F=1.05$. In Fig.~\ref{F.profiles}b one can appreciate that the profile becomes 
increasingly kink-shaped, so that, in the thermodynamic limit, the chain is split into two 
symmetric regions, each one characterized by a rotation frequency equal to the
value imposed at the boundary ($F_1$ and $F_N$, respectively). The two regions are separated
by a localized interfacial area, where $T_n$ is finite and $\omega_n$
changes from $F_1$ to $F_N$.

The supercritical phase is, however, more complex than revealed by this average characterization. 
In Fig.~\ref{F.profiles}c we plot a space-time representation of the ``instantaneous"
frequency profile $\Omega_n = \langle p_n \rangle_\tau$, where the average is performed over
a time $\tau = 10^4$, that is much longer than the microscopic time scale and  significantly 
shorter than the slow hydrodynamic scales. This representation reveals that the transition 
region is quite thin and fluctuates. Data has been reported for $F=1.3$ and $N=6400$
to show that this behavior is robust also for large values of the torque and for long chains.
 A more quantitative analysis can be performed by 
studying the shape of the instantaneous frequency profile. In practice we  study
$\Delta\Omega_m=\langle\Omega_{m+\hat n(t)+1}-\Omega_{m+\hat n(t)}\rangle$, 
where $m$ is the distance from the instantaneous position $\hat n(t)$ of the temperature peak.
The data
in Fig.~\ref{F.profiles}d corresponds to different values of $F$ and is plotted after rescaling $\Delta\Omega_m$ and $m$ by 
$N^{1/5}$ and $N^{-1/5}$, respectively. The good data collapse of the curves corresponding
to various system sizes reveals the presence of a second scaling exponent.

The overall scenario can be described in the following way. On the one hand,
the $N^{-1/2}$ scaling of the average profiles is related to the decay rate of the 
strength of the effective force which pins the interfacial region in the middle of the
chain. On the other hand, the $N^{-1/5}$ scaling of the width of the instantaneous 
active region is related to the maintenance of the momentum flux, which, in fact, 
scales with the same exponent, $j^p \sim N^{-1/5}$ (see the inset in Fig.~\ref{F.profiles}).

In order to further refine our understanding of the supercritical regime, 
we have computed the spectra of Lyapunov exponents $\lambda_n$ ($n=1,\ldots ,2N$). 
As a first check we have verified that the sum of all $\lambda_n$ is equal to the dissipation 
$-2\gamma$, as it should. The spectra obtained for different system sizes reveal
substantial differences from the standard (extensive) chaotic regime \cite{Ruelle1982,Livi1986}.
For increasing $N$, most exponents decrease and approach zero, indicating a weakly
chaotic behaviour, consistently with the zero-temperature imposed at the boundaries.
The spectrum does not, however, uniformly shrinks to zero
as $\lambda_1$ and $\lambda_{2N}$ remain finite. In the upper inset of Fig.~\ref{F.lyspectr},
one can see that the largest exponent $\lambda_1(N)$ approaches a constant $\Lambda = 0.262$
up to corrections of order $N^{-2/3}$. The existence of finite exponents is consistent 
with the observation of a finite temperature in the central region of the lattice
where some chaotic dynamics persists in the thermodynamic limit.

A further inspection of Fig.~\ref{F.lyspectr} also reveals that the Lyapunov spectra cross the zero
axis at a finite value $n_u/N\approx 0.6$, indicating that the dimension density of the 
unstable manifold is finite, i.e. this observable is extensive.
The same is true also for the Kaplan-Yorke (KY) dimension, that increases with $N$ and 
possibly converges to 2, meaning that the non-equilibrium invariant measure extends
along (almost) directions. Surprisingly, a qualitatively different behavior is observed 
for KS  entropy-density $h_{KS}$, estimated as the area under the positive
part of the Lyapunov spectrum.  The results for different system sizes are 
reported in the lower inset of Fig.~\ref{F.lyspectr}. There, we see that $h_{KS}$
vanishes, revealing a non-extensive nature of the chaotic dynamics. By extrapolating
from the largest simulations, one can conjecture a decay as $N^{-1/2}$.
A yet more detailed analysis could be performed by investigating the convergence of
the bulk of the Lyapunov spectrum, but this task would require considering much
larger systems and we leave it to future studies.

The above features are rather unusual with respect to usual space-time chaotic
system whereby dynamical invariants are extensive with the volume \cite{Ruelle1982,Livi1986}.
They are instead partially reminiscent of delayed dynamical systems, where, for large
delays, the KY dimension is extensive (i.e. proportional to
the delay), while the KS entropy remains finite \cite{DoyneFarmer1982,Lepri1994}. 
Here, however, this is a true instance of space-time chaos and the non-extensive character 
of the KS entropy is not a formal consequence of the interpretation of the delay 
as a spatial extension.

Additional studies carried out for larger $F$ values confirm the general validity of 
this mixed extensive/non-extensive behavior. One has only to be careful in selecting
sufficiently long chains, so as to avoid the existence of a point-like interfacial region:
this phenomenon, which is suggestive of a second transition
(see the open circles in Fig.~\ref{F.phasediag}), is instead a finite size effect 
that disappears for sufficiently large values of $N$ (see the diamonds in Fig.~\ref{F.phasediag}).
 
{\bf Finite temperature}. We now explore the behavior for non-zero boundary temperatures 
(i.e. in the presence of an external source of noise).
The results reported in Fig.~\ref{F.phasediag} indicate that for $T=0.5$ (crosses and plusses) 
both $\hat T$ and the momentum flux depend smoothly on $F$. Additionally, the momentum 
conductivity is normal, i.e $j^p \sim 1/N$. For smaller temperatures, a residue of the 
transition is still present as a sudden increase of $\hat T$ at finite $F$ 
(see Fig.~\ref{F.phasediag}a  for $T=0.1$). This is, nevertheless, a finite-size effect, 
as the pseudo-discontinuity disappears upon increasing $N$ 
(compare squares and triangles in Fig.~\ref{F.phasediag}a).
This suggests that the fluctuations imposed on the boundaries induce phase--slips which
are thereby responsible for the suppression of the synchronized state in the subcritical
region. It is however interesting to notice the persistence of a bump in the temperature 
profile, although its width is now proportional to $N$.

{\bf DNLS model}. The above discussed non--equilibrium transition is not
a peculiarity of the rotor model. Here below, we show that a similar scenario can be
observed for the DNLS equation,
$i \dot {z}_n = -2 |z_n|^2z_n - z_{n+1}-z_{n-1}$, where $z_n = (p_n + i q_n)/\sqrt{2} $ is a
complex variable.
The DNLS Hamiltonian has two conserved quantities, the mass/norm $a$ and the energy 
density $h$ \cite{Rasmussen2000,Iubini2013}, so that it is a natural candidate for describing
coupled transport \cite{Iubini2012,Iubini2013a}.   

\begin{figure}[ht]
\begin{center}
\includegraphics[width=0.4\textwidth,clip]{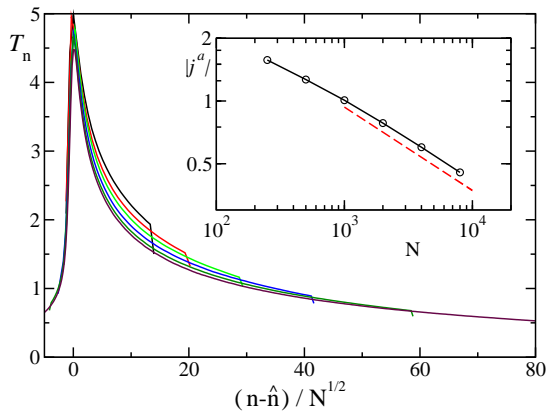}
\caption{(Color online) Kinetic temperature profiles of the DNLS equation for  $N=250$, 
500, 1000, 2000, 4000, 8000 and  $T=0$, $\mu_1=2$ and $\mu_N=5$.
The spatial direction is rescaled in order to obtain a data collapse in the region around the point
of maximal temperature $\hat n$. Inset: chemical-potential flux  $j_n^a$  versus system size $N$. 
The (red) dashed line corresponds to a slope $-2/5$.}
\label{F.DNLS}
\end{center}
\end{figure}

We have numerically studied a DNLS chain interacting with two Langevin thermostats 
at $T=0$ and different chemical potentials $\mu_1$ and $\mu_N$ imposed 
at the boundaries (see Ref.~\cite{Iubini2013a} for details). In this case, 
the control parameter, i.e. the driving force, is $\delta\mu = |\mu_N-\mu_1|/2$ 
\cite{Iubini2013a}. When $\delta\mu$ is larger than a critical value, 
(e.g., $\mu_1 = 2$ and $\mu_N=5$), a bumpy temperature profile spontaneously emerges. 
In Fig.~\ref{F.DNLS} one can see that the width of the peak scales as $N^{1/2}$, 
(as for the XY chain), while the left-right symmetry is lost and the mass (norm) flux 
$j_n^a=i\langle(z_nz_{n-1}^*-z_n^*z_{n-1})\rangle$ now scales
as $N^{-2/5}$ instead of $N^{-1/5}$ as in the XY case (see the inset in Fig.~\ref{F.DNLS}). 

{\bf Discussion and conclusions}.
In this Letter we have shown that in the presence of coupled transport, the application of
deterministic boundary forces may induce a non equilibrium transition at zero
temperature. 

The different scaling behavior observed in two models (XY and DNLS), suggests the
existence of multiple universality classes.  We conjecture that, as
in the context of (anomalous) heat conduction in one-dimensional systems, the
discriminating factor is given by the presence of symmetries~\cite{Delfini07b,Spohn2013}. 
In the XY case, the average value of the torque is immaterial, since it can be 
removed by selecting a suitably rotating frame. This invariance implies that positive 
and negative frequency shifts are equivalent to one another and, as a result, 
symmetric profiles are expected and, indeed, observed. 
This symmetry is, however, not present in the DNLS dynamics, in spite of the fact 
that the DNLS equation can be effectively approximated by an XY model~\cite{Iubini2013a} in
the limit of small gradients and large mass densities.

If finite-temperature heat baths are considered, the first-order transition is smoothed
out and the anomalous superconductive behavior is replaced by a normal transport.
A characterization of such a phase remains, however, nontrivial because of the
underlying kink in the frequency/chemical potential profile, which induces a
bump in the temperature and  makes the implementation of a Green-Kubo formalism 
rather problematic.
It will be instructive to explore the same problem in higher dimensions: 
one cannot exclude that the zero-temperature transition reported herein survives 
in the presence of finite fluctuations. 

It is finally important to stress the anomalous properties of the supercritical phase in
the context of nonlinear dynamics and synchronization phenomena. This regime is 
indeed characterized by a mixture of extensive (Kaplan-Yorke dimension) and 
non-extensive (Kolmogorov-Sinai entropy) properties, which make it atypical and 
different from: (i) the standard extensive chaos typically observed 
in both dissipative and Hamiltonian models; (ii)
the localized chaotic states generated when all oscillators are damped
and driven \cite{Bonart1999,Martinez1999}. In other words, this regime provides
an example of how complex dynamics can be in relatively simple
models, characterized by nearest neighbour interactions.


\end{document}